\documentclass[preprint,pra]{revtex4-1}

\usepackage{graphicx}
\usepackage{dcolumn}
\usepackage{bm}
\usepackage{amsmath}
\usepackage{amssymb}
\usepackage{latexsym}
\usepackage{epsfig}
\usepackage{amsbsy}
\usepackage{array}
\usepackage{setspace}
\usepackage{mathrsfs}
\usepackage{geometry}
\usepackage{cancel}
\begin{document}
\thispagestyle{empty}

\title{The NRQED Hamiltonian and photon-exchange interaction up to $m\alpha^8$ order}

\author{Wanping Zhou}
\affiliation{School of Physics and Technology, Wuhan University, Wuhan 430000 China}

\affiliation{Engineering and Technology College, Hubei University of Technology, Wuhan 430000 China}
\author{Xuesong Mei}
\affiliation{School of Physics and Technology, Wuhan University, Wuhan 430000 China}
\author{Haoxue Qiao \footnote{Haoxue Qiao; electronic mail: qhx@whu.edu.cn}}
\affiliation{School of Physics and Technology, Wuhan University, Wuhan 430000 China}

\begin{abstract}
 
We derive the effective Hamiltonian of the Nonrelativistic Quantum Electrodynamic up to $m\alpha^8$ by using scattering matching approach. At $m\alpha^6$ order, these results are coincide with Pachucki's, which is obtained by applying Foldy-Wouthuysen transformation. And by using the NRQED Hamiltonian, we derive the photon-exchange interaction in non-retarded approximation and the retardation correction up to $m\alpha^8$. The energy shift of the photon-exchange interaction is obtained by studying the pole of the total Green function.

\end{abstract}

\maketitle

\newpage

 \section{INTRODUCTION}
The non-relativistic few-body systems, such as Hydrogen, Helium, Hydrogen molecule ions and lithium, are interesting both in experimental and theoretical fields. As the energy-level of these systems can be calculated and measured in high-precision. It is always used to  determine unit of time \citep{PhysRevLett.113.023004} and the physical constant such as fine-structure constant $\alpha$ and Rydberg constant \citep{PhysRevLett.118.063001,RevModPhys.80.633} or test the quantum electrodynamic theory.

The energy-level of non-relativistic atoms and molecules $E$ can be expanded in power series of the fine-structure constant $\alpha$
   \begin{equation}
  \begin{aligned}
   E=&E_{(2)}+\Delta E_{(4)}+\Delta E_{(5)}
   +\Delta E_{(6)}+\Delta E_{(7)}+\Delta E_{(8)}+....
   \end{aligned} 
  \end{equation}
where the subscript $(n)$ means the $m\alpha^n$. The leading terms $E_{(2)}$, which is calculated by solving the Schr$\ddot{o}$dinger equation, is at $m\alpha^2$ order. The high-order energy shift, relativistic, radiation and recoil corrections, are the means value of the high-order Hamiltonian. The precision of the energy is restricted by the accuracy of $E_{(2)}$, the non-relativistic wave-functions and the order of relativistic, radiation and recoil corrections obtained. As the non-relativistic problem has been solved by using Rayleigh-Ritz variational method in Hylleraas coordinates \citep{JPhysB.39.2095,PhysRevA.98.012510,arXiv:math-ph/0605018,PhysRevA.83.034503}. The core issue of the high-precision calculating the energy-level is deriving these high-order corrections. 

The most effective method to do this work is the non-relativistic quantum electrodynamic(NRQED) \citep{CASWELL1986437,PhysRevD.53.4909,PhysRevD.55.7267}, which is an effective field theory describing the non-relativistic electromagnetic systems. The key point is the contributions of the virtual photon are separated into the high-energy region $\omega>m\alpha$ and low-energy region $\omega\backsim m\alpha^{2}$. In the low-energy region, the energy shift can be calculated by using the low-energy effective Hamiltonian. In the High-energy region, the contributions can be treated in QED approach with scattering approximation. 

The NRQED has achieved a great success in the light Hydrogenlike atoms. Their energy-level has been calculated up to $m\alpha^7$ \citep{IEIDES200163} In the three-body systems, Helium and Hydrogen molecule ions, the effective Hamiltonian of the energy up to $m\alpha^6$ order has been complete obtained \citep{PhysRevA.71.012503} And at $m\alpha^7$ order, only the fine-structure corrections are derived \citep{PhysRevLett.104.070403}. The studying about the $m\alpha^8$ order corrections are seldom reported.     

The fine-structure splitting in the $2^{3}P_{J}$ level of Helium is recognized as the best atomic system for the determination of the fine-structure constant $\alpha$. Comparing with calculations of the $m\alpha^7$ order energy, the measured result can be used to determine the fine-structure constant with a precision of about $2\times 10^{-9}$ \citep{PhysRevLett.118.063001} It is necessary to derive the $m\alpha^8$ order Hamiltonian to reduce the uncertain $2.3\times 10^{-10}$ in CODATA \citep{RevModPhys.88.035009} 

At present, the low-energy effective Hamiltonian of NRQED obtained by using Foldy-Wouthuysen transformation \citep{PhysRevA.82.052520} is at $m\alpha^6$ order. The $m\alpha^8$ order result hasn't been obtained. In this work, we study the effective Hamiltonian of NRQED up to $m\alpha^8$ by using the scattering matching \citep{CASWELL1986437,PhysRevD.53.4909,PhysRevD.55.7267}. And it will cross check the result up to $m\alpha^6$ order. The second purpose of the paper is deriving the photon-exchange interaction by using this effective Hamiltonian of NRQED. It is parts of contribution to the $m\alpha^8$ order energy. These results can be used to calculate the relativistic, recoil and radiation corrections \citep{PhysRevA.74.022512,PhysRevD.53.4909,PhysRevA.82.052520,PhysRevA.71.012503,JPhysB.31.5123,CanJPhys.77.267,PhysRevD.58.093013,PhysRevA.95.062510,PhysRevA.94.052508,PhysRevA.95.012508}.

In next section, firstly, we calculate one-photon-fermion scattering amplitude on-shell and find their contribution to the effective Hamiltonian. Secondly, we calculate the two-photon-fermion scattering amplitude on-shell by separating the contribution of positive-energy and negative-energy mid-states. After subtracting the pole term, which is the contributions of Hamiltonian induced by one-photon-fermion scattering, we can obtain the two-photon-fermion contact terms and the effective Hamiltonian of NRQED up to $m\alpha^8$ order. Then, we discuss the arbitrariness of choosing the one-photon-fermion vertex and give the general formula of the effective Hamiltonian of NRQED up to $m\alpha^8$ order. This arbitrariness will change the two-photon-fermion vertex, and the effective Hamiltonian will have infinite equivalent formulas. In the third section, We study photon-exchange interaction up to $m\alpha^8$ by using this effective Hamiltonian of NRQED. The instantaneous parts and the retarded parts of the photon-exchange interactions are derived separately. The final section is the summation.

 \section{THE NRQED HAMILTONIAN}
 
The NRQED is equivalent to the QED as one could obtain the same scattering amplitude (scattering matching) \citep{CASWELL1986437,PhysRevD.53.4909,PhysRevD.55.7267}. However, the NRQED Hamiltonian should contain more and more terms in order to improving precision. Generally, one could neglect the higher-order terms, which are irrelevant to the accuracy by using magnitude order estimation: Every term in the NRQED Hamiltonin consists of $\partial _{t}$, the electron momentum $\vec{\pi}$ (or $\vec{p}$, $\vec{\partial}$), electromagnetic $\vec{E},\vec{B}$ (or$A^{\mu}$) and Pauli matrices $\sigma^{i}$ It is well known that their magnitudes in light atoms are $\langle p \rangle \simeq \langle \vec{\partial} \rangle \simeq O(m\alpha)$, $\langle \partial _{t} \rangle \simeq \langle mv^{2} \rangle \simeq  O(m\alpha^{2})$, $\langle eA^{0} \rangle \simeq O(m\alpha^{2})$, $\langle eA^{i} \rangle \simeq O(m\alpha^{3})$, $\langle eE \rangle \simeq O(m^{2}\alpha^{3})$, $\langle eB \rangle \simeq O(m^{2}\alpha^{4})$ \citep{PhysRevD.53.4909}. As an example, we can neglect the $B^{2}\pi^{2}$ terms at $m\alpha^{8}$ order.

   The NRQED Hamiltonian up to $m\alpha^6$ order \citep{PhysRevA.82.052520,PhysRevA.71.012503} is widely used in the atomic energy-level calculation. It is
    \begin{equation}\label{HFW6}
  \begin{aligned}
    \mathscr{H}=& 
     \dfrac{1}{2}\tilde{\pi}^{2}+\varphi
    -\dfrac{1}{8}\tilde{\pi}^{4}
    -\dfrac{i}{8}[\tilde{\pi},\tilde{E}]
     +\dfrac{1}{16}\tilde{\pi}^{6}
      +\dfrac{3i}{64}
     \{\tilde{\pi}^{2},[\tilde{\pi},\tilde{E}]\}
      \\&
     +\dfrac{5xi}{128}
     [\tilde{\pi}^{2},\{\tilde{\pi},\tilde{E}\}]
     -\dfrac{5(1-x)i}{64}\{\tilde{\pi},\partial_{t}\tilde{E}\}
     +\dfrac{5x-1}{32}\tilde{E}^{2}      
     , 
   \end{aligned} 
  \end{equation}
where the tilde means $\tilde{f}=\sigma^{i}f^{i}$, $x$ is arbitrary constant in virtue of  unitary transformation (we choose $\epsilon^{ij}$ in Ref. \citep{PhysRevA.82.052520} be proportional to $\propto\delta^{ij}$), $\kappa$ is the anomalous magnetic moment and we will always omit the mass and charge of the fermion without confusion. However, this result is only obtained by using Foldy-Wouthuysen (FW) transformation and hasn't been obtained by using the scattering matching procedures. We will prove the result is the same at $m\alpha^6$ order, and derive the effective NRQED Hamiltonian up to $m\alpha^8$ order. 
 
  \subsection{The one-photon-fermion scattering matching}
 The one-photon-fermion interaction can be described by the Hamiltonian in QED,   
 \begin{equation}
    \mathscr{H}_{int}=\bar{\psi}(p')
    (\gamma^{\mu}+\dfrac{i}{2}\kappa\sigma^{\mu\nu}q_{\nu})
    eA_{\mu}(q)\psi(p),
   \end{equation}  
where $e$,$\kappa$ are charge and anomalous magnetic moment of the fermion. 

The Dirac fermion 
\[ u(p)=\sqrt{\dfrac{E+m}{2E}}
  \left( 
    \begin{aligned}
      1_{2\times 2}\phi   \\ 
      \dfrac{\sigma\cdot\vec{p}\phi}{E+m}
    \end{aligned}    
  \right),
\] 
is on-shell $E^{2}=\vec{p}^{2}+m^{2}$, and $\phi$ is a 2 component non-relativistic fermion, which will be omit in the non-relativistic expansion for simplicity in this paper. In other words, all the no relativistic scattering amplitude $(...)$ is $\phi^{\dagger}(...)\phi$ in this work.
 
 The non-relativistic expansion of Coulomb-photon-fermion scattering amplitude is 
  \begin{equation}\label{current0}
  \begin{aligned}
   J^{0}\equiv &\bar{u}(p')\gamma^{0}u(p)
   \\=&
   1+\dfrac{5x}{128}(\textbf{p}'^{2}-\textbf{p}^{2})^{2}
   +\dfrac{5(1-x)}{64}\omega(\textbf{p}'^{2}-\textbf{p}^{2})
   -\dfrac{14+5x}{512} (\textbf{p}'^{2}+\textbf{p}^{2})
   (\textbf{p}'^{2}-\textbf{p}^{2})^{2}  
   \\&
  +\dfrac{1}{8}\left(
  1-\dfrac{3}{8}(\textbf{p}'^{2}+\textbf{p}^{2})
  +\dfrac{40\textbf{p}'^{2}+40\textbf{p}^{2}
  -9(\textbf{p}'^{2}-\textbf{p}^{2})^{2}}{128}
   \right) 
   \left(
    2\tilde{\textbf{p}}'\tilde{\textbf{p}}
   -\textbf{p}'^{2}-\textbf{p}^{2}
   \right) 
   \\&+o(m\alpha^{8}).
   \end{aligned}
  \end{equation}
The $x$ is arbitrary constant. Because the energy of the photon satisfy conservation of energy $\omega=E'-E\simeq \dfrac{\vec{p}'^{2}}{2}-\dfrac{\vec{p}^{2}}{2}+...$. There are infinite ways to expand the amplitude. This is the origin, where the arbitrary constant $x$ in Eq.(\ref{HFW6}) come from. It is not hard to write down the effective Hamiltonian corresponding to the Coulomb-photon-fermion scattering amplitude.
  \begin{equation}\label{H_CPF}
  \begin{aligned}
    \mathscr{H}_{CPF}=& 
     \varphi
     +\dfrac{5xi}{128}
     [\tilde{\pi}^{2},\{\tilde{\pi},\tilde{E}\}]
     -\dfrac{5(1-x)i}{64}\{\tilde{\pi},\partial_{t}\tilde{E}\}
     -\dfrac{(14+5x)i}{512}
     [\tilde{\pi}^{4},\{\tilde{\pi},\tilde{E}\}]
     \\&
      -\dfrac{i}{8}[\tilde{\pi},\tilde{E}]
      +\dfrac{3i}{64}
     \{\tilde{\pi}^{2},[\tilde{\pi},\tilde{E}]\}
     -\dfrac{40i}{1024}
     \{\tilde{\pi}^{4},[\tilde{\pi},\tilde{E}]\}
     +\dfrac{9i}{1024}
     [\tilde{\pi}^{2},[\tilde{\pi}^{2},[\tilde{\pi},\tilde{E}]]] 
     \\&+o(m\alpha^{8}).
   \end{aligned} 
  \end{equation}

The nonrelativistic expansion of transverse-photon-fermion scattering amplitude is
   \begin{equation}\label{currenti}
  \begin{aligned}
   J^{i}\equiv&\bar{u}(p')\gamma^{i}u(p)
   \\=&
   \dfrac{1}{2}\left(
   1-\dfrac{1}{4}(\textbf{p}'^{2}+\textbf{p}^{2})
   +\dfrac{1}{8}
   (\textbf{p}'^{4}+\textbf{p}^{4}
   +\textbf{p}'^{2}\textbf{p}^{2})
   \right)
   (\tilde{\textbf{p}}'\sigma^{i}+\sigma^{i}\tilde{\textbf{p}}) 
   \\&   
   +\dfrac{1}{8}
   \left( 1-\dfrac{3}{8}(\textbf{p}'^{2}+\textbf{p}^{2}))\right) 
   \omega
   (\tilde{\textbf{p}}'\sigma^{i}-\sigma^{i}\tilde{\textbf{p}})
   \\&
   +\dfrac{5}{128}\omega
   \left( x(\textbf{p}'^{2}-\textbf{p}^{2})+2(1-x)\omega\right) 
   (\tilde{\textbf{p}}'\sigma^{i}+\sigma^{i}\tilde{\textbf{p}})
   +o(m\alpha^{8}),
   \end{aligned}
  \end{equation}
where the second and third line have a factor $\omega$, these terms come from the terms containing $\vec{E}=-\partial_{t}\vec{A}+...$ in the $\mathscr{H}_{CPF}$, and the first line in the transverse-photon-fermion scattering indicates the Hamiltonian 
 \begin{equation}
  \begin{aligned}
    \mathscr{H}_{TPF}=& 
    \dfrac{1}{2}\{\tilde{\textbf{p}},\tilde{\textbf{A}}\}
   -\dfrac{1}{8}
   \{\tilde{\textbf{p}}^{2},
   \{\tilde{\textbf{p}},\tilde{\textbf{A}}\}\}
   +\dfrac{1}{16}
   \left(
    \tilde{\textbf{p}}^{4}
   \{\tilde{\textbf{p}},\tilde{\textbf{A}}\}+
   \{\tilde{\textbf{p}},\tilde{\textbf{A}}\}
   \tilde{\textbf{p}}^{4}+
   \tilde{\textbf{p}}^{2}
   \{\tilde{\textbf{p}},\tilde{\textbf{A}}\}
   \tilde{\textbf{p}}^{2}\right), 
   \end{aligned} 
  \end{equation}  
they are parts of the relativistic kinetic energy expansion $\dfrac{1}{2}\tilde{\pi}^{2}-\dfrac{1}{8}\tilde{\pi}^{4}+\dfrac{1}{16}\tilde{\pi}^{6}+...$. 
 
The anomalous magnetic moment part of one-photon-fermion scattering amplitude can be obtained by using the similar way. They are
 \begin{equation}
  \begin{aligned}
 \dfrac{i}{2}(q^{2})\bar{u}(p')\sigma^{0\nu}q_{\nu}u(p)=&
 -\dfrac{1}{4}
 (\tilde{\textbf{p}}'\tilde{\textbf{q}}-
  \tilde{\textbf{q}}\tilde{\textbf{p}}) 
  +\dfrac{1}{16}(\textbf{p}'^{2}+\textbf{p}^{2})
 (\tilde{\textbf{p}}'\tilde{\textbf{q}}-
  \tilde{\textbf{q}}\tilde{\textbf{p}})
  \\&
  +\dfrac{1}{32}(\textbf{p}'^{2}-\textbf{p}^{2})
 (\tilde{\textbf{p}}'\tilde{\textbf{q}}+
  \tilde{\textbf{q}}\tilde{\textbf{p}}),
   \end{aligned} 
  \end{equation}
and  
 \begin{equation}
  \begin{aligned}
 \dfrac{i}{2}\bar{u}(p')\sigma^{i\nu}q_{\nu}u(p)=&
  \dfrac{1}{4}
   (\sigma^{i}\tilde{\textbf{q}}-\tilde{\textbf{q}}\sigma^{i})
  +\dfrac{1}{16}\tilde{\textbf{p}}'
   (\sigma^{i}\tilde{\textbf{q}}- \tilde{\textbf{q}}\sigma^{i})
   \tilde{\textbf{p}}
  -\dfrac{1}{32}(\textbf{p}'^{2}+\textbf{p}^{2})
  (\sigma^{i}\tilde{\textbf{q}}- \tilde{\textbf{q}}\sigma^{i})
     \\&
  -\dfrac{\omega}{4}
  (\sigma^{i}\tilde{\textbf{p}}-
  \tilde{\textbf{p}}'\sigma^{i}).
     \end{aligned} 
  \end{equation}
 The effective Hamiltonians corresponding to these terms are  
 \begin{equation}\label{H_APF}
  \begin{aligned}
    \mathscr{H}_{APF}=& 
     -\dfrac{i\kappa}{4}[\tilde{\pi},\tilde{E}]
     +\dfrac{i\kappa}{16}
     \{\tilde{\pi}^{2},[\tilde{\pi},\tilde{E}]\}
     +\dfrac{i\kappa}{32}
     [\tilde{\pi}^{2},\{\tilde{\pi},\tilde{E}\}]
     -\dfrac{\kappa}{2}\tilde{B}
     +\dfrac{\kappa}{16}[\tilde{\pi},[\tilde{\pi},\tilde{B}]].
   \end{aligned} 
  \end{equation}
  
By using the method in the section, one can derive the effective Hamiltonian which contain only one $E$ or $B$ field operator. The term contains N $E$ or $B$ field operators should be studied the N-photon-fermion scattering. As the $\langle E \rangle =O(m^{2}\alpha^3)$,$\langle B \rangle =O(m^{2}\alpha^4)$, we need to derive the $EEpp$, $BB$ and $EBp$ terms in NRQED Hamiltonian to achieve $m\alpha^8$ order. In order to obtain these terms, the two-photon-fermion scattering matching, which is the Compton scattering, should be studied.

\subsection{The two-photon-fermion scattering matching}
 The amplitude of two-photon-fermion scattering is 
 \begin{equation}
  \begin{aligned}
  \mathscr{M}&=J^{\mu\nu}A_{\mu}(q)A_{\nu}(q')
  \\&=
  \bar{u}(p')\left( 
  \gamma^{\nu}\dfrac{1}{\cancel{p}+\cancel{q}-m}\gamma^{\mu}
  +
  \gamma^{\mu}\dfrac{1}{\cancel{p}+\cancel{q}'-m}\gamma^{\nu}
  \right)u(p)A_{\mu}(q)A_{\nu}(q'),
      \end{aligned} 
  \end{equation}
where $p'-p=q+q'$. This amplitude has poles corresponding to the relativistic kinetic energy. The pole terms should be the product of the one-photon-fermion vertices and the propagator, as the NRQED and QED will have the same result as the energy of the mid-state approaching the pole. Subtracting the pole terms from the amplitude of two-photon-fermion scattering, The difference indicates two-photon-fermion contact terms should be included in the NRQED Hamiltonian. We try to obtain the two-photon-fermion contact term by setting $x=1$ for simplicity.

Using the positive/negative energy projection operator $\Lambda_{\pm}=\dfrac{E_{\textbf{p}}\gamma^{0}\mp\textbf{p}\cdot\gamma+m}{ 2E_{\textbf{p}}}$, the two-photon-fermion scattering amplitude can be rewritten as

   \begin{equation}
  \begin{aligned}
  J^{\mu\nu}=J^{\mu\nu}_{+}+J^{\mu\nu}_{-},
  \end{aligned} 
  \end{equation}
  
 \begin{equation}
  \begin{aligned}
  J^{\mu\nu}_{+}=&
  \bar{u}(p')\left( 
  \dfrac{\gamma^{\nu}\Lambda_{+}(\textbf{p+q})\gamma^{\mu}}
  {E_{\textbf{p}}+\omega-E_{\textbf{p+q}}}
  + 
  \dfrac{\gamma^{\mu}\Lambda_{+}(\textbf{p}+\textbf{q}')\gamma^{\nu}}
  {E_{\textbf{p}}+\omega'-E_{\textbf{p}+\textbf{q}'}}
  \right)u(p),
      \end{aligned} 
  \end{equation}

 \begin{equation}
  \begin{aligned}
  J^{\mu\nu}_{-}=&
   \bar{u}(p')\left( 
   \dfrac{\gamma^{\nu}\Lambda_{-}(\textbf{p+q})\gamma^{\mu}}
  {E_{\textbf{p}}+\omega+E_{\textbf{p}+\textbf{q}}}
  +
   \dfrac{\gamma^{\mu}\Lambda_{-}(\textbf{p}+\textbf{q}')\gamma^{\nu}}
  {E_{\textbf{p}}+\omega'+E_{\textbf{p+q'}}}
  \right)u(p),
      \end{aligned} 
  \end{equation}
where $E_{\textbf{p}}=\sqrt{\textbf{p}^{2}+m^{2}}$, and $J^{\mu\nu}_{\pm}$ are the positive and negative mid-state's contributions to the two-photon-fermion scattering.
 
The positive energy part $J^{\mu\nu}_{+}$ has poles at $E_{\textbf{p}}+\omega'=E_{\textbf{p}+\textbf{q}'}$ or $E_{\textbf{p}}+\omega=E_{\textbf{p+q}}$. However, The numerators of the positive energy part aren't the product of two one-photon-fermion vertex, which is obtained in the previous subsection. We define currents 
$J^{\mu}(p',p)\equiv\bar{u}(p')\gamma^{\mu}u(p)=
J^{\mu}_{0}(\textbf{p}',\textbf{p})+
\omega J^{\mu}_{1}(\textbf{p}',\textbf{p})+...$
, which is the Taylor expansion in power series of  $\omega$ in the Sec.II. $J^{\mu}_{0}$ and $J^{\mu}_{1}$ are zeroth order and first-order term in the Taylor expansion at $\omega=0$.

The positive energy part $J^{\mu\nu}_{+}$ is
   \begin{equation}
  \begin{aligned}
 J^{\mu\nu}_{+}=&
  =\dfrac{J^{\nu}(p',p'-q')J^{\mu}(p+q,p)}
  {E_{\textbf{p}}+\omega-E_{\textbf{p+q}}}
  + 
  \dfrac{J^{\mu}(p',p'-q)J^{\nu}(p+q',p)}
  {E_{\textbf{p}}+\omega'-E_{\textbf{p}+\textbf{q}'}}
  \\&
  -J^{\nu}_{0}(\textbf{p}',\textbf{p}'-\textbf{q}')
  J^{\mu}_{1}(\textbf{p+q},\textbf{p})
  +J^{\nu}_{1}(\textbf{p}',\textbf{p}'-\textbf{q}')
  J^{\mu}_{0}(\textbf{p+q},\textbf{p})
  \\&
  -J^{\mu}_{0}(\textbf{p}',\textbf{p}'-\textbf{q})
  J^{\nu}_{1}(\textbf{p}+\textbf{q}',\textbf{p})
  +J^{\mu}_{1}(\textbf{p}',\textbf{p}'-\textbf{q})
  J^{\nu}_{0}(\textbf{p}+\textbf{q}',\textbf{p})+o(m\alpha^8).
      \end{aligned} 
  \end{equation}  
The second and third lines are parts of contact terms. As these terms haven't had any pole, 

The negative energy part has no pole, They are other parts of the  contact terms. We can expand it straight
 \begin{equation}
  \begin{aligned}
 J^{\mu\nu}_{-}=&
   \dfrac{J^{\nu}_{\phi\chi}(p',p'-q')J^{\mu}_{\chi\phi}(p+q,p)}
  {E_{\textbf{p}}+\omega+E_{\textbf{p}+\textbf{q}}}
  +
   \dfrac{J^{\mu}_{\phi\chi}(p',p'-q)J^{\nu}_{\chi\phi}(p+q',p)}
  {E_{\textbf{p}}+\omega'+E_{\textbf{p+q'}}}
  \\=&
  \dfrac{1}{2}(1-\dfrac{\omega-\omega'}{4})
  J^{\nu}_{\phi\chi}J^{\mu}_{\chi\phi}
  +\dfrac{1}{2}(1-\dfrac{\omega'-\omega}{4})
  J^{\mu}_{\phi\chi}J^{\nu}_{\chi\phi}
  \\&
  -\dfrac{1}{16}\left( 
  \textbf{p}'^{2}J^{\nu}_{\phi\chi}J^{\mu}_{\chi\phi}+
  J^{\nu}_{\phi\chi}(\textbf{p}+\textbf{q})^{2}
  J^{\mu}_{\chi\phi}+
  J^{\nu}_{\phi\chi}J^{\mu}_{\chi\phi}\textbf{p}^{2}
  \right)+
  \\&
  -\dfrac{1}{16}\left( 
  \textbf{p}'^{2}J^{\mu}_{\phi\chi}J^{\nu}_{\chi\phi}+
  J^{\mu}_{\phi\chi}(\textbf{p}+\textbf{q}')^{2}
  J^{\nu}_{\chi\phi}+
  J^{\mu}_{\phi\chi}J^{\nu}_{\chi\phi}\textbf{p}^{2}
  \right)+o(m\alpha^8), 
      \end{aligned} 
  \end{equation}
where the current $J_{\phi\chi},J_{\chi\phi}$ are the one-photon-fermion-antifermion scattering amplitudes
 \begin{equation}
  \begin{aligned}
   J^{0}_{\phi\chi}=&\bar{u}(p')\gamma^{0}v(p)   
   =-J^{0}_{\chi\phi}=-\bar{v}(p')\gamma^{0}u(p)
   \\=&
   \dfrac{1}{2}
   \left( 1-\dfrac{1}{4}(\textbf{p}'^{2}+\textbf{p}^{2})\right) 
    \tilde{\textbf{q}}
    +\dfrac{1}{2}
     \left( \textbf{p} + \textbf{p}'\right)\cdot\textbf{q}
    \left( \tilde{\textbf{p}} + \tilde{\textbf{p}}'\right) ,
   \end{aligned}
  \end{equation}
and  
    \begin{equation}
  \begin{aligned}
   J^{i}_{\phi\chi}=&\bar{u}(p')\gamma^{i}v(p)   
   =J^{i}_{\chi\phi}=\bar{v}(p')\gamma^{i}u(p)
   \\=&
   \sigma^{i}
   -\dfrac{1}{4}\tilde{\textbf{p}}'\sigma^{i}\tilde{\textbf{p}}
   -\dfrac{1}{8}(\textbf{p}'^{2}+\textbf{p}^{2})\sigma^{i}.
   \end{aligned}
  \end{equation}
$v$ is the plane wave function of antifermion.

The total contact terms of total two-photon-fermion scattering scattering amplitude are
   \begin{equation}
  \begin{aligned}
  J^{\mu\nu}_{contact}=&
  \dfrac{1}{2}(1-\dfrac{\omega-\omega'}{4})
  J^{\nu}_{\phi\chi}J^{\mu}_{\chi\phi}
  +\dfrac{1}{2}(1-\dfrac{\omega'-\omega}{4})
  J^{\mu}_{\phi\chi}J^{\nu}_{\chi\phi}
  \\&
  -\dfrac{1}{16}\left( 
  \textbf{p}'^{2}J^{\nu}_{\phi\chi}J^{\mu}_{\chi\phi}+
  J^{\nu}_{\phi\chi}(\textbf{p}+\textbf{q})^{2}
  J^{\mu}_{\chi\phi}+
  J^{\nu}_{\phi\chi}J^{\mu}_{\chi\phi}\textbf{p}^{2}
  \right)+
  \\&
  -\dfrac{1}{16}\left( 
  \textbf{p}'^{2}J^{\mu}_{\phi\chi}J^{\nu}_{\chi\phi}+
  J^{\mu}_{\phi\chi}(\textbf{p}+\textbf{q}')^{2}
  J^{\nu}_{\chi\phi}+
  J^{\mu}_{\phi\chi}J^{\nu}_{\chi\phi}\textbf{p}^{2}
  \right)+
  \\&
  -J^{\nu}_{0}(\textbf{p}',\textbf{p}'-\textbf{q}')
  J^{\mu}_{1}(\textbf{p+q},\textbf{p})
  +J^{\nu}_{1}(\textbf{p}',\textbf{p}'-\textbf{q}')
  J^{\mu}_{0}(\textbf{p+q},\textbf{p})
  \\&
  -J^{\mu}_{0}(\textbf{p}',\textbf{p}'-\textbf{q})
  J^{\nu}_{1}(\textbf{p}+\textbf{q}',\textbf{p})
  +J^{\mu}_{1}(\textbf{p}',\textbf{p}'-\textbf{q})
  J^{\nu}_{0}(\textbf{p}+\textbf{q}',\textbf{p})+... 
      \end{aligned}. 
  \end{equation} 
After tedious and trivial deducing, the $00$ component indicates the NRQED Hamiltonian containing double $E$ terms is
  \begin{equation}
  \begin{aligned}
    \mathscr{H}_{00C}=&
     \dfrac{1}{8}\tilde{E}^{2} 
     -\dfrac{1}{64}
     \{\tilde{E},\{\tilde{\pi},\{\tilde{\pi},\tilde{E}\}\}\}
     -\dfrac{3}{64}
     \{\tilde{E},\{\tilde{\pi}^{2},\tilde{E}\}\}
     -\dfrac{i}{16}[\tilde{E},\partial_{t}\tilde{E}].
   \end{aligned} 
  \end{equation}
The contribution of the contact term with $0i$ and $ij$ should be contained in the NRQED Hamiltonian in Sec.II for self-consistency. It is easy to check that is true. The contribution of anomalous magnetic moment part is $\dfrac{1}{8}(2\kappa+\kappa^{2})\tilde{E}^{2} $, which can be obtained by using the same method. 

The total Hamiltonian of NRQED up to $m\alpha^8$ order in this work is 
 \begin{equation}\label{Hamlitonian} 
  \begin{aligned}
    \mathscr{H}=& 
     \dfrac{1}{2}\tilde{\pi}^{2}+\varphi
    -\dfrac{1}{8}\tilde{\pi}^{4}
    -\dfrac{i}{8}[\tilde{\pi},\tilde{E}]
     +\dfrac{1}{16}\tilde{\pi}^{6}
      +\dfrac{3i}{64}
     \{\tilde{\pi}^{2},[\tilde{\pi},\tilde{E}]\}
     +\dfrac{5i}{128}
     [\tilde{\pi}^{2},\{\tilde{\pi},\tilde{E}\}]
     +\dfrac{1}{8}\tilde{E}^{2} 
      \\&
     -\dfrac{5}{128}\tilde{\pi}^{8}
     -\dfrac{19i}{512}
     [\tilde{\pi}^{4},\{\tilde{\pi},\tilde{E}\}]
     -\dfrac{40i}{1024}
     \{\tilde{\pi}^{4},[\tilde{\pi},\tilde{E}]\}
     +\dfrac{9i}{1024}
     [\tilde{\pi}^{2},[\tilde{\pi}^{2},[\tilde{\pi},\tilde{E}]]] 
     \\&
     -\dfrac{1}{64}
     \{\tilde{E},\{\tilde{\pi},\{\tilde{\pi},\tilde{E}\}\}\}
     -\dfrac{3}{64}
     \{\tilde{E},\{\tilde{\pi}^{2},\tilde{E}\}\}
     -\dfrac{i}{16}[\tilde{E},\partial_{t}\tilde{E}]
     \\&
     -\dfrac{i\kappa}{4}[\tilde{\pi},\tilde{E}]
     +\dfrac{i\kappa}{16}
     \{\tilde{\pi}^{2},[\tilde{\pi},\tilde{E}]\}
     +\dfrac{i\kappa}{32}
     [\tilde{\pi}^{2},\{\tilde{\pi},\tilde{E}\}]
     -\dfrac{\kappa}{2}\tilde{B}
     +\dfrac{\kappa}{16}[\tilde{\pi},[\tilde{\pi},\tilde{B}]]
     \\&
     +\dfrac{1}{8}(2\kappa+\kappa^{2})\tilde{E}^{2}.
   \end{aligned} 
  \end{equation}
The first line is the same as the result of Pachucki \citep{PhysRevA.82.052520,PhysRevA.71.012503}, and we have fixed $x=1$. The second and third lines are at $m\alpha^{8}$ order, which are obtained by one photon scattering matching and two photon scattering matching. The last two lines are contributions of anomalous magnetic moment. 

\subsection{The one-photon-fermion vertex and the two-photon-fermion vertex}

The different $x$ in the one-photon-fermion vertex will bring additional terms to the two-photon-fermion vertex. In this subsection, we will obtain the additional terms and all the equivalent results of the Hamiltonian. 
  
As the arbitrariness of choosing $x$ originate from the energy conservation $\omega=E'-E$. The one-photon-fermion vertex can be written as $V=[M,H]+N$ or $V'=-\omega M+N$, where the $H$ is the Hamiltonian of the free fermion, and the operators $M,N$ will be given later. The contributions of $V$ and $V'$ are the same in the one-photon-fermion scattering, the differences in the two-photon-fermion scattering are the additional terms to the two-photon-fermion contact term

It is easy to check the equation  
  \begin{equation}
  \begin{aligned}
 &\phi^{\dagger}(E')
 \left( V^{\dagger}(\omega')\dfrac{1}{E-H+\omega}V(\omega)+
 V(\omega)\dfrac{1}{E-H-\omega'}V^{\dagger}(\omega')\right) 
 \phi(E)=
 \\&  
 \phi^{\dagger}(E')
 \left( V'^{\dagger}(\omega')\dfrac{1}{E-H+\omega}V'(\omega)+
 V'(\omega)\dfrac{1}{E-H-\omega'}V'^{\dagger}(\omega')+
 W_{2-photon}\right) 
 \phi(E),
   \end{aligned} 
  \end{equation}
where  
  \begin{equation}
  \begin{aligned}
 W_{2-photon}=\dfrac{1}{2}\left( 
 [M,[M^{\dagger},H]]+[M^{\dagger},[M,H]]+(\omega'+\omega)[M,M^{\dagger}]
 \right)+[M^{\dagger},N]+[N^{\dagger},M].
   \end{aligned} 
  \end{equation}
It means the additional terms compensate for the deviation brought by the transformation $V\rightarrow V'$, with the requirement of the invariant of the scattering amplitude As the scattering amplitude is the same, the equivalent effective Hamiltonian can be obtained by adding the $-V+V'+W_{2-photon}$ term.  
The first two terms transform the one photon term $V$ of the old effective Hamiltonian into one photon term $V'$, The $W_{2-photon}$ terms compensate for the deviation. 

As the operator $M=-i\{\tilde{p},\tilde{E}\}+...$ and $N=\varphi-\dfrac{i}{8}[\tilde{\pi},\tilde{E}]+...$  in one photon vertex are at $m^{3}\alpha^4$ and $m\alpha^2$ order. In the $W_{2-photon}$, the two photon terms in the parentheses are at $m\alpha^{10}$ order, which is negligible. Most important terms are
  \begin{equation}
  \begin{aligned}
 W_{2-photon}&=[M^{\dagger},N]+[N^{\dagger},M]+o(m\alpha^{10})
 \\&
 =\left[ i\{\tilde{p},\tilde{E^{\dagger}}\},\varphi-\dfrac{i}{8}[\tilde{p},\tilde{E}]\right]+
 \left[\varphi^{\dagger}-\dfrac{i}{8}[\tilde{p},\tilde{E^{\dagger}}], -i\{\tilde{p},\tilde{E}\}\right]+o(m\alpha^{10})
  \\&
 =-2\{\tilde{E},\tilde{E^{\dagger}}\}
+\dfrac{1}{8}
\left[\{\tilde{p},\tilde{E^{\dagger}}\},[\tilde{p},\tilde{E}]\right]+
\dfrac{1}{8}
\left[\{\tilde{p},\tilde{E}\},[\tilde{p},\tilde{E^{\dagger}}]\right]
 +o(m\alpha^{10}).
   \end{aligned} 
  \end{equation}
  
The equivalent transformation $V\rightarrow V'$ and the additional terms imply the effective Hamiltonian in the last subsection can be added the following term.
  \begin{equation}
  \begin{aligned}
 \mathscr{H}'=&-\dfrac{5(1-x)}{64} \bigg\{
  i[\dfrac{\tilde{\pi}^{2}}{2},\{\tilde{\pi},\tilde{E}\}]
  +\{\tilde{\pi},\partial_{t}\tilde{E}\}+\{\tilde{E},\tilde{E}\}
  \\&
  -i[\dfrac{\tilde{\pi}^{4}}{8},\{\tilde{\pi},\tilde{E}\}]
  -\dfrac{1}{8}
  \left[\{\tilde{\pi},\tilde{E}\},[\tilde{\pi},\tilde{E}]\right]
  \bigg\}.
   \end{aligned} 
  \end{equation}
The first two term is $V'-V$ at $m\alpha^6$ order. The third term is the additional two photon contact term $W_{2-photon}$ at $m\alpha^6$ order. The first three terms have been obtained by using FW transformation \citep{PhysRevA.82.052520}. The last two terms are $V'-V$ and $W_{2-photon}$ terms at $m\alpha^8$ order. They are obtained first time.

\section{THE PHOTON-EXCHANGE INTERACTION BETWEEN FERMION} 
As $\phi,\varphi,A$ in NRQED Hamiltonian are the operators of the fields, The NRQED Hamiltonian can't be used to calculate the energy shift in the bound state directly. The photon-exchange interactions of the fermion should be studied first. These photon-exchange interactions have corrections to the total Green function. The energy shift can be derived by calculating the pole of the total Green function.

The one photon-exchange interaction between two fermions a and b in the bound state is
 \begin{equation}
  \begin{aligned}
  \Sigma_{1p}&=
  \sum_{a\neq b}\int\dfrac{d^{4}k}{i(2\pi)^{4}}
  G_{\mu\nu} (k)
  J^{\mu}_{a}(k)e^{i\vec{k}\cdot\vec{r}_{a}}
  G(E_{0}-k^{0})
  J^{\nu}_{b}(-k)e^{-i\vec{k}\cdot\vec{r}_{b}}
  \\&
  +\sum_{a\neq b}\int\dfrac{d^{4}k}{i(2\pi)^{4}}
  G_{\mu\nu} (k)
  J^{\mu}_{a}(k)e^{i\vec{k}\cdot\vec{r}_{a}}
  G(E_{0}-k^{0})T_{(4)}G(E_{0}-k^{0})
  J^{\nu}_{b}(-k)e^{-i\vec{k}\cdot\vec{r}_{b}}+o(m\alpha^{8}),
   \end{aligned} 
  \end{equation}
where $J^{\mu}_{a}$ is the one-photon-fermion scattering amplitude Eq.(\ref{current0})(\ref{currenti}) in the Sec.II. The second line is first-order perturbation of the relativistic kinetic energy  $T_{(4)}=-\dfrac{1}{8}\sum p^{4}_{a}$. Because we find this term is at $m\alpha^7$ order, the higher-order perturbations of relativistic kinetic energy are neglectable up to $m\alpha^8$ order. The fermion Green function is defined as $G(E)=(E_{0}-H_{0}+i0^{+})^{-1}$, and the propagators of photon $G_{\mu\nu}$ in Coulomb gauge are 
   \begin{equation}
  \begin{aligned}
   G_{\mu\nu} (k)=\Bigg\{   \begin{array}{cc}
   \dfrac{-1}{\vec{k}^{2}},&\mu=\nu=0\\
   \dfrac{-d^{ij}(\vec{k})}{k_{0}^{2}-\vec{k}^{2}+i0^{+}}, &\mu=i,\nu=j
   \end{array}    
      \end{aligned} 
  \end{equation}
where $d^{ij}(\vec{k})=\left(\delta_{ij}-\dfrac{k_{i}k_{j}}{\vec{k}^{2}} \right)$. 

The two photon-exchange interactions between fermions in the bound system are   
  \begin{equation}
  \begin{aligned}
  \Sigma_{2p}&=\sum_{a,b,c,d,unequal}\int
  \dfrac{G_{\mu\nu} (k)d^{4}k}{i(2\pi)^{4}}
  \dfrac{G_{\alpha\beta} (q)d^{4}q}{i(2\pi)^{4}}\times
    \\& 
  \Big[
  J^{\beta}_{c}(q)e^{i\vec{q}\cdot\vec{r}_{c}}G(E_{0}-q^{0})    
  J^{\mu\alpha}_{a}(k,-q)e^{i(\vec{k}-\vec{q})\cdot\vec{r}_{a}}
  G(E_{0}-k^{0})J^{\nu}_{b}(-k)e^{-i\vec{k}\cdot\vec{r}_{b}}+
    \\& 
   J^{\nu}_{b}(k)e^{i\vec{k}\cdot\vec{r}_{b}}G(E_{0}-k^{0})
  J^{\beta}_{c}(q)e^{i\vec{q}\cdot\vec{r}_{c}}G(E_{0}-k^{0}-q^{0})    
  J^{\mu\alpha}_{a}(-k,-q)e^{-i(\vec{k}+\vec{q})\cdot\vec{r}_{a}}+
  \\&
  J^{\mu\alpha}_{a}(k,q)e^{i(\vec{k}+\vec{q})\cdot\vec{r}_{a}}
  G(E_{0}-k^{0}-q^{0})J^{\nu}_{b}(-k)e^{-i\vec{k}\cdot\vec{r}_{b}}
  G(E_{0}-q^{0}) J^{\beta}_{c}(-q)e^{-i\vec{q}\cdot\vec{r}_{c}}+
  \\&
  J^{\mu}_{a}(k)e^{i\vec{k}\cdot\vec{r}_{a}}G(E_{0}-k^{0})
  J^{\alpha}_{b}(q)e^{i\vec{q}\cdot\vec{r}_{a}}G(E_{0}-k^{0}-q^{0})
  J^{\nu}_{c}(-k)e^{-i\vec{k}\cdot\vec{r}_{c}}G(E_{0}-q^{0})
  J^{\beta}_{d}(-q)e^{-i\vec{q}\cdot\vec{r}_{d}}+
  \\&
  J^{\mu}_{a}(k)e^{i\vec{k}\cdot\vec{r}_{a}} G(E_{0}-k^{0})
  J^{\alpha}_{b}(q)e^{i\vec{q}\cdot\vec{r}_{a}}G(E_{0}-k^{0}-q^{0})
  J^{\beta}_{c}(-q)e^{-i\vec{q}\cdot\vec{r}_{c}} G(E_{0}-k^{0})
  J^{\nu}_{d}(-k)e^{-i\vec{k}\cdot\vec{r}_{d}}
  \Big]
  ,
   \end{aligned} 
  \end{equation}
where the $J^{\mu\nu}$ is the contact term in the Sec.II. The three photon-exchange interaction is neglectable up to $m\alpha^8$, as the leading order of three photon-exchange potentials, such as $E^{3},A^{3}$, are at $m\alpha^9$ order. 

In this section, we will study the instantaneous and the retarded interactions separately, 
 \begin{equation}
  \begin{aligned}
  \Sigma_{1p}=\Sigma_{1pi}+\Sigma_{1pr},
   \end{aligned} 
  \end{equation}
where the subscript $i,r$ is the abbreviation of instantaneous and retarded. As there are two exchange-photon propagators in $\Sigma_{2p}$, the two exchange-photon interaction is the sum
  \begin{equation}
  \begin{aligned}
  \Sigma_{2p}=\Sigma_{2pii}+\Sigma_{2pir}+\Sigma_{2prr}.
   \end{aligned} 
  \end{equation} 
The instantaneous interactions are $\Sigma_{1pi}$ and $\Sigma_{2pii}$. We will derive them firstly.

\subsection{The interaction between fermions in nonretarded approximation}

The interaction between fermions in nonretarded approximation can be obtained by neglecting the $k^{0}$ in the $G_{\mu\nu}$. And there are two key points to derive the instantaneous and the retarded interactions repeatedly. Firstly, as the product  $J^{\mu}_{a}(k)J^{\nu}_{b}(k^{0})$ or $J^{\mu}_{a}(k)J^{\nu\lambda}_{b}(k^{0})$ contains at most a single power of $k^{0}$ in our work, the integral of $q^0$ or $k^0$ can be done by using the symmetrization $(q^{0}\leftrightarrow-q^{0})$ and the equalities
 \begin{equation}\label{q0integral}
  \begin{aligned}
  &\dfrac{1}{2}\int^{\infty}_{-\infty}\dfrac{dq^{0}}{i(2\pi)}
  \Big(\dfrac{1}{A-q^{0}+i0^{+}}
  +\dfrac{1}{A+q^{0}+i0^{+}}\Big)=-\dfrac{1}{2};
  \\&
  \dfrac{1}{2}\int^{\infty}_{-\infty}\dfrac{dq^{0}}{i(2\pi)}
  \Big(\dfrac{1}{A-q^{0}+i0^{+}}\dfrac{1}{B-q^{0}+i0^{+}}
  +(q^{0}\leftrightarrow-q^{0})\Big)=0;
  \\&
  \dfrac{1}{2}\int^{\infty}_{-\infty}\dfrac{dq^{0}}{i(2\pi)}
  \Big(\dfrac{1}{A-q^{0}+i0^{+}}\dfrac{1}{B-q^{0}+i0^{+}}
  \dfrac{1}{c-q^{0}+i0^{+}}
  +(q^{0}\leftrightarrow-q^{0})\Big)=0.
   \end{aligned} 
  \end{equation}   
Secondly, the Coulomb photon-exchange part is always instantaneous, because the $G_{00}$ is independent of $k^0$ in Coulomb gauge. 

Using the Eq.(\ref{q0integral}), we obtain these photon-exchange interactions. The instantaneous part of one photon-exchange interaction is 
   \begin{equation}
  \begin{aligned}
  \Sigma_{1pi}&=-\dfrac{1}{2}\sum_{a\neq b}\int 
  \dfrac{d^{3}k}{(2\pi)^{3}}G'_{\mu\nu}(\vec{k})
  J^{\mu}_{a}(\vec{k})e^{i\vec{k}\cdot\vec{r}_{a}}
  J^{\nu}_{b}(-\vec{k})e^{-i\vec{k}\cdot\vec{r}_{b}}
  =\dfrac{1}{2}\sum_{a}J^{\mu}_{a}(-i\vec{\partial})
  \mathscr{A}_{\mu}(\vec{r}_{a}),
   \end{aligned} 
  \end{equation}
and the instantaneous part of two photon-exchange interaction is
  \begin{equation}
  \begin{aligned}
  \Sigma_{2pii}
  =&\sum_{a,b,c,unequal}\int
  \dfrac{d^{3}k}{(2\pi)^{3}}G'_{\mu\nu}(\vec{k})
  \dfrac{d^{3}q}{(2\pi)^{3}}G'_{\alpha\beta}(\vec{q})
  J^{\beta}_{c}(q)e^{i\vec{q}\cdot\vec{r}_{c}}   
  J^{\mu\alpha}_{a}(k,-q)e^{i(\vec{k}-\vec{q})\cdot\vec{r}_{a}}
  J^{\nu}_{b}(-k)e^{-i\vec{k}\cdot\vec{r}_{b}}
  \\&
  =\dfrac{1}{4}\sum_{a}
  \mathscr{A}_{\alpha}(\vec{r}_{a})   
  J^{\mu\alpha}_{a}(-i\overrightarrow{\partial},i\overleftarrow{\partial})
  \mathscr{A}_{\mu}(\vec{r}_{a}),
   \end{aligned} 
  \end{equation}
where $G_{\mu\nu}(\vec{k})\equiv G_{\mu\nu}(\vec{k},k^{0}=0)$, the right/left arrow in $J^{\mu\alpha}_{a}$ means the partial acts on the right/left electromagnetic field $\mathscr{A}$, and the electromagnetic field 4-vector $\mathscr{A}$ produced by other particles is 
 \begin{equation}
  \begin{aligned}
  \mathscr{A}_{\mu}(\vec{r}_{a})=-\sum_{a\neq b} 
  \int\dfrac{d^{3}x_{b}}{(2\pi)^{3}}
  G_{\mu\nu} (\vec{r}_{ab})J^{\nu}(\vec{r}_{b}),
   \end{aligned} 
  \end{equation}
where propagators of photon in non-retarded approximation are   
  \begin{equation}
  \begin{aligned}
   G_{\mu\nu} (\vec{r})=\int \dfrac{d^{3}k}{(2\pi)^{3}} G_{\mu\nu} (\vec{k},k^{0}=0)e^{i\vec{k}\cdot \vec{r}}
   =\dfrac{1}{4\pi}\Bigg\{   
   \begin{array}{cc}
   -\dfrac{1}{r},&\mu=\nu=0\\
   \dfrac{1}{2r}\left(\delta_{ij}+\dfrac{r_{i}r_{j}}{r^{2}} \right). &\mu=i,\nu=j
   \end{array}    
      \end{aligned} 
  \end{equation} 

As the photon-exchange interactions in non-retarded approximation are the product of the current and the electromagnetic potential, $\Sigma_{1pi}$ and $\Sigma_{2pii}$ can be expanded by using the non-relativistic expansions of the electromagnetic field 4-vector and the current.
  
The non-relativistic expansions of the electromagnetic field 4-vector are $\varphi=\varphi_{(2)}+\varphi_{(4)}+\varphi_{(6)}...$ and $\mathscr{A}^{i}=\mathscr{A}^{i}_{(3)}+\mathscr{A}^{i}_{(5)}+\mathscr{A}^{i}_{(7)}+...$, where the subscript $(n)$ means the order is at $m\alpha^n$. Substituting the currents, which are one-photon-fermion scattering amplitudes Eq.(\ref{current0})(\ref{currenti}), into $\mathscr{A}_{\mu}(x)$, these terms in the expansions of the scalar potential are 
  \begin{equation}
  \begin{aligned}
   \varphi_{(2)}(\vec{r}_{a} )=&-\sum_{b\neq a}G_{00}(\vec{r}_{ab}), 
   \\ 
   \varphi_{(4)}(\vec{r}_{a} )=&\sum_{b\neq a} 
   \dfrac{1}{8}[\tilde{p}_{b},-i\tilde{\partial}_{b}G_{00}(\vec{r}_{ab})], 
   \\ 
   \varphi_{(6)}(\vec{r}_{a} )=&-\sum_{b\neq a} 
   \left( \dfrac{3}{64}{p_{b}^{2},[\tilde{p}_{b},-i\tilde{\partial}_{b}G_{00}(\vec{r}_{ab})]}
   +\dfrac{5}{128}[p_{b}^{2},[p_{b}^{2},G_{00}(\vec{r}_{ab})]]\right),
   \\ 
   \varphi_{(8)}(\vec{r}_{a} )=&\sum_{b\neq a}\Big(
    \dfrac{38}{1024}\{p_{b}^{2},[p_{b}^{2},[p_{b}^{2},G_{00}(\vec{r}_{ab})]]\}+
   \dfrac{40}{1024}\{p_{b}^{4},[\tilde{p}_{b},-i\tilde{\partial}_{b}G_{00}(\vec{r}_{ab})]\}
   \\&
   -\dfrac{9}{1024}[p_{b}^{2},[p_{b}^{2},[\tilde{p}_{b},-i\tilde{\partial}_{b}G_{00}(\vec{r}_{ab})]]] \Big). 
      \end{aligned} 
  \end{equation}
Those terms in the expansions of the vector potential are   
  \begin{equation}
  \begin{aligned}
   \mathscr{A}^{i}_{(3)}(\vec{r}_{a} )=&
   \dfrac{1}{2}\sum_{b\neq a} \{\tilde{p}_{b},\sigma^{j}_{b}G^{ij}(\vec{r}_{ab})\},
   \\ 
   \mathscr{A}^{i}_{(5)}(\vec{r}_{a} )=&
   -\dfrac{1}{8}\sum_{b\neq a}\{p_{b}^{2},\{\tilde{p}_{b},\sigma^{j}_{b}G^{ij}(\vec{r}_{ab})\}\}, 
  \\ 
   \mathscr{A}^{i}_{(7)}(\vec{r}_{a} )=&
   \dfrac{1}{8}\sum_{b\neq a}\Big(
   \{p_{b}^{4},\{\tilde{p}_{b},\sigma^{j}_{b}G^{ij}(\vec{r}_{ab})\}\}
   +p_{b}^{2}\{\tilde{p}_{b},\sigma^{j}_{b}G^{ij}(\vec{r}_{ab})\}p_{b}^{2}\Big).   
      \end{aligned} 
  \end{equation}
It is obvious that the $\varphi_{(2)}$ and $\mathscr{A}_{(3)}$ are the Coulomb potential and non-relativistic vector potential, and the higher-order terms are relativistic corrections to the electromagnetic field 4-vector.
 
Substituting the non-retarded electromagnetic 4-vector into 
$\Sigma_{1pi},\Sigma_{2pii}$, we obtain the Hamiltonian in non-retarded approximation, which is the sum of the kinetic energy $T$ and instantaneous interaction term $V$. The kinetic energy terms are 
  \begin{equation}\label{kineticenergy}
  \begin{aligned}
   T=&\sum_{a}\left( \dfrac{1}{2}p^{2}_{a}-\dfrac{1}{8}p^{4}_{a}
   +\dfrac{1}{16}p^{6}_{a}-\dfrac{5}{128}p^{8}_{a}+o(m\alpha^8)\right).    
   \end{aligned} 
  \end{equation}
    The instantaneous interaction terms are
  \begin{equation}\label{potentialenergy}
  \begin{aligned}
   V=V_{(2)}+V_{(4)}+V_{(6)}+V_{(8)}++o(m\alpha^8),    
   \end{aligned} 
  \end{equation}
where    
   \begin{equation}\label{relativisticpotential}
  \begin{aligned}
   V_{(2)}=&\sum_{a}\varphi_{a(2)}, 
   \\ 
   V_{(4)}=&\sum_{a}\left(
   -\dfrac{1}{2}\{\tilde{p}_{a},\tilde{\mathscr{A}}_{a(3)}\}+ 
   \varphi_{a(4)}
   -\dfrac{i}{8}[\tilde{p}_{a},\tilde{\mathscr{E}}_{a(3)}]\right) , 
   \\ 
   V_{(6)}=&\sum_{a}\Big(
   -\dfrac{1}{2}\{\tilde{p}_{a},\tilde{\mathscr{A}}_{a(5)}\}
   +\dfrac{1}{2}\mathscr{A}^{2}_{a(3)} 
   +\varphi_{a(6)}
   +\dfrac{1}{8}\{\tilde{\mathscr{A}}_{a(3)},\tilde{p}_{a}p^{2}_{a}\}
   \\&
   +\dfrac{1}{8}\tilde{p}_{a}\{\tilde{\mathscr{A}}_{a(3)},\tilde{p}_{a}\}\tilde{p}_{a}
   -\dfrac{i}{8}[\tilde{p}_{a},\tilde{\mathscr{E}}_{a(5)}]
   +\dfrac{i}{8}[\tilde{\mathscr{A}}_{a(3)},\tilde{\mathscr{E}}_{a(3)}]
   \\&
      +\dfrac{3i}{64}\{p^{2}_{a},[\tilde{p}_{a},\tilde{\mathscr{E}}_{a(3)}]\}
     +\dfrac{5i}{128}
     [p^{2}_{a},\{\tilde{p}_{a},\tilde{\mathscr{E}}_{a(3)}\}]
     +\dfrac{1}{8}\mathscr{E}^{2}_{a(3)} 
   \Big),
   \\ 
   V_{(8)}=&\sum_{a}\Big[
   -\dfrac{1}{2}\{\tilde{p}_{a},\tilde{\mathscr{A}}_{a(7)}\}
   +\mathscr{A}_{a(3)}\cdot \mathscr{A}_{a(5)} 
   +\varphi_{a(8)}
   +\dfrac{1}{8}\{\tilde{\mathscr{A}}_{a(5)},\tilde{p}_{a}p^{2}_{a}\}
   \\&
   +\dfrac{1}{8}\tilde{p}_{a}\{\tilde{\mathscr{A}}_{a(5)},\tilde{p}_{a}\}\tilde{p}_{a}
   -\dfrac{1}{8}\{\mathscr{A}^{2}_{a(3)},p^{2}_{a}\} 
   -\dfrac{1}{8}\{\tilde{\mathscr{A}}_{a(3)}\tilde{p}_{a}\tilde{\mathscr{A}}_{a(3)},\tilde{p}_{a}\}
   \\& 
   -\dfrac{1}{8}\tilde{\mathscr{A}}_{a(3)}p_{a}^{2}\tilde{\mathscr{A}}_{a(3)} 
   -\dfrac{1}{8}\tilde{p}_{a}\mathscr{A}_{a(3)}^{2}\tilde{p}_{a}  
   -\dfrac{i}{8}[\tilde{p}_{a},\tilde{\mathscr{E}}_{a(7)}]
   +\dfrac{i}{8}[\tilde{\mathscr{A}}_{a(3)},\tilde{\mathscr{E}}_{a(5)}]
   \\&
   -\dfrac{1}{16}\Big(
   \{\tilde{\mathscr{A}}_{a(3)},\tilde{p}_{a}p^{4}_{a}\}
   +\tilde{p}_{a}\{\tilde{\mathscr{A}}_{a(3)},\tilde{p}_{a}p^{2}_{a}\}\tilde{p}_{a} 
   +p_{a}^{2}\{\tilde{\mathscr{A}}_{a(3)},\tilde{p}_{a}\}p_{a}^{2}
    \Big)    
   \\&
      +\dfrac{3i}{64}\Big(
      \{p^{2}_{a},[\tilde{p}_{a},\tilde{\mathscr{E}}_{a(5)}]\}
      -
      \{p^{2}_{a},[\tilde{\mathscr{A}}_{a(3)},\tilde{\mathscr{E}}_{a(3)}]\}
      -
      \{\{\tilde{p}_{a},\tilde{\mathscr{A}}_{a(3)}\},[\tilde{p}_{a},\tilde{\mathscr{E}}_{a(3)}]\}
      \Big)
   \\&
     +\dfrac{5i}{128}\Big(
     [p^{2}_{a},\{\tilde{p}_{a},\tilde{\mathscr{E}}_{a(5)}\}]
     -[p^{2}_{a},\{\tilde{\mathscr{A}}_{a(3)},\tilde{\mathscr{E}}_{a(3)}\}]
     -[\{\tilde{p}_{a},\tilde{\mathscr{A}}_{a(3)}\},\{\tilde{p}_{a},\tilde{\mathscr{E}}_{a(5)}\}]
     \Big)
   \\&
     +\dfrac{1}{4}\mathscr{E}_{a(3)}\cdot \mathscr{E}_{a(5)}
   \\&
     -\dfrac{38i}{1024}
     [p^{4}_{a},\{\tilde{p}_{a},\tilde{\mathscr{E}}_{a(3)}\}]
     -\dfrac{40i}{1024}
     \{p^{4}_{a},[\tilde{p}_{a},\tilde{\mathscr{E}}_{a(3)}]\}
     +\dfrac{9i}{1024}
     [p^{2}_{a},[\tilde{p}^{2}_{a},[\tilde{p}_{a},\tilde{\mathscr{E}}_{a(3)}]]]
    \\&
      -\dfrac{1}{64}
     \{\tilde{\mathscr{E}}_{a(3)},\{\tilde{p}_{a},\{\tilde{p}_{a},\tilde{\mathscr{E}}_{a(3)}\}\}\}
     -\dfrac{3}{64}
     \{\tilde{\mathscr{E}}_{a(3)},\{\tilde{p}^{2}_{a},\tilde{\mathscr{E}}_{a(3)}\}\}  
   \Big], 
   \end{aligned} 
  \end{equation}
the electric field $\vec{\mathscr{E}}_{(n+1)}=-\nabla\varphi_{(n)}$. The $V_{(2)}$ is the Coulomb interaction, $V_{(4)}$ is the leading order term of the Breit interaction, and $V_{(6)}, V_{(8)}$ are the $m\alpha^6, m\alpha^8$ order terms of the Breit interaction. Most of the terms in  $V_{(6)}$ have been obtained by Pachucki in Ref.\citep{PhysRevA.71.012503}. The $m\alpha^8$ order terms $V_{(8)}$ have never been obtained before. 

The Hamiltonian of the bound system in non-retarded approximation is
    \begin{equation}\label{Hnr}
    H=T+V=H_{0}+H_{(4)}+H_{(6)}+H_{(8)}+....
    \end{equation}   
where the $H_{0}$ is the non-relativistic leading term, and the relativistic parts $H_{(n)}=T_{(n)}+V_{(n)}$.

 As an example, in the Hydrogen-like atoms $\varphi_{(2)}=-\dfrac{1}{r}$, $\mathscr{E}_{(3)}=\dfrac{\sigma\cdot\vec{r}}{r^{3}}$, $\varphi_{(4)}=\varphi_{(6)}=\varphi_{(8)}=0$ and $\mathscr{A}^{i}_{(3)}=\mathscr{A}^{i}_{(5)}=\mathscr{A}^{i}_{(7)}=0$, the Hamiltonian of Hydrogen-like atom in non-retarded approximation is 
 \begin{equation}
  \begin{aligned}
 H=&\dfrac{1}{2}p^{2}+\varphi_{(2)}
  -\dfrac{1}{8}p^{4}
  -\dfrac{i}{8}[\tilde{p},\tilde{\mathscr{E}}_{(3)}]
   +\dfrac{1}{16}p^{6}
  +\dfrac{3i}{64}\{p^{2},[\tilde{p},\tilde{\mathscr{E}}_{(3)}]\}
     +\dfrac{5i}{128}
     [p^{2},\{\tilde{p},\tilde{\mathscr{E}}_{(3)}\}]
     +\dfrac{1}{8}\mathscr{E}^{2}_{(3)},
   \\&
     -\dfrac{5}{128}p^{8}  
     -\dfrac{38i}{1024}
     [p^{4},\{\tilde{p},\tilde{\mathscr{E}}_{(3)}\}]
     -\dfrac{40i}{1024}
     \{p^{4},[\tilde{p},\tilde{\mathscr{E}}_{(3)}]\}
     +\dfrac{9i}{1024}
     [p^{2},[\tilde{p}^{2},[\tilde{p},\tilde{\mathscr{E}}_{(3)}]]]
    \\&
      -\dfrac{1}{64}
     \{\tilde{\mathscr{E}}_{(3)},\{\tilde{p},\{\tilde{p},\tilde{\mathscr{E}}_{(3)}\}\}\}
     -\dfrac{3}{64}
     \{\tilde{\mathscr{E}}_{(3)},\{\tilde{p}^{2},\tilde{\mathscr{E}}_{(3)}\}\}+o(m\alpha^8). 
   \end{aligned} 
  \end{equation}

\subsection{The retarded corrections to energy} 

The retarded corrections can be obtained by subtracting the instantaneous parts $\Sigma_{1i}$ and $\Sigma_{2ii}$ from the photon-exchange interactions $\Sigma_{1p}$ and $\Sigma_{2p}$. The retarded part of one photon-exchange interaction is
 \begin{equation}
  \begin{aligned}
  \Sigma_{1pr}
  =&\sum_{a\neq b}\int 
  \dfrac{d_{ij}(\vec{k})d^{3}k}{2\omega(2\pi)^{3}}
  J^{i}_{a}(k)e^{i\vec{k}\cdot\vec{r}_{a}}
  \tilde{G}(E_{0},\omega)
  J^{j}_{b}(-k)e^{-i\vec{k}\cdot\vec{r}_{b}}
  \\
  +&\sum_{a\neq b}\int
  \dfrac{d_{ij}(k)d^{3}k}{2\omega_{k}(2\pi)^{3}}
  J^{i}_{a}(k,q)e^{i\vec{k}\cdot\vec{r}_{a}}
  G(E_{0}-\omega_{k})T_{(4)}G(E_{0}-\omega_{k})
  J^{j}_{b}(-k)e^{-i\vec{k}\cdot\vec{r}_{b}}+o(m\alpha^8),
   \end{aligned} 
  \end{equation}    
  $\omega=\left|\vec{k}\right|$ and the retarded Green function $\tilde{G}$ is defined as 
    \begin{equation}
    \tilde{G}(E_{0},\omega)\equiv\dfrac{E_{0}-H_{0}}{\omega(E_{0}-H_{0}-\omega+i0^{+})}.
    \end{equation}   

The one retarded photon part of two-photon-exchange interaction is 
 \begin{equation}
  \begin{aligned}
  \Sigma_{2pir}
  =&\sum_{a\neq c}\int
  \dfrac{d_{ij}(k)d^{3}k}{2\omega_{k}(2\pi)^{3}}
  \Big\{
  J^{i}_{a}(k,q)e^{i\vec{k}\cdot\vec{r}_{a}}
  G(E_{0}-\omega_{k})V_{(4)}
  G(E_{0}-\omega_{k})
  J^{j}_{c}(-k)e^{-i\vec{k}\cdot\vec{r}_{c}}+
  \\&
  J^{i}_{ex,a}(k)e^{i\vec{k}\cdot\vec{r}_{a}} 
    \tilde{G}(E_{0},\omega_{k})
  J^{j}_{c}(-k)e^{-i\vec{k}\cdot\vec{r}_{c}}+
  J^{j}_{c}(k)e^{i\vec{k}\cdot\vec{r}_{c}}
    \tilde{G}(E_{0},\omega_{k}) 
    J^{i}_{ex,a}(-k)e^{-i\vec{k}\cdot\vec{r}_{a}}
    \Big\}+o(m\alpha^8),
   \end{aligned} 
  \end{equation}
where the current $J^{i}_{ex,a}(k)$ is induced by the other fermion b
 \begin{equation}
  \begin{aligned}
   J^{i}_{ex,a}(k)=&-2\sum_{b\neq a}\int \dfrac{d^{3}q}{2\omega_{q}^{2}(2\pi)^{3}}G'_{\mu\nu}(\vec{k})
  J^{\mu}_{b}(q)e^{i\vec{q}\cdot\vec{r}_{ab}}    
  J^{i\nu}_{a}(k,-q),
   \end{aligned} 
  \end{equation} 
  
The $\Sigma_{1pr}$ and $\Sigma_{2pir}$ have one retarded photon-exchange interaction. The total contribution of one retarded photon is defined as
  \begin{equation}
  \begin{aligned}
  \Sigma_{1r}(E_{0})=&\Sigma_{1pr}+\Sigma_{2pir}
  \\&
  =\sum_{a\neq b}\int 
  \dfrac{d_{ij}(\vec{k})d^{3}k}{2\omega(2\pi)^{3}}
  \Big\{J^{i}_{T,a}(k)e^{i\vec{k}\cdot\vec{r}_{a}}
  \tilde{G}(E_{0},\omega)
  J^{j}_{T,b}(-k)e^{-i\vec{k}\cdot\vec{r}_{b}}
  \\&
  +J^{i}_{a}(k,q)e^{i\vec{k}\cdot\vec{r}_{a}}
  G(E_{0}-\omega_{k})H_{(4)}G(E_{0}-\omega_{k})
  J^{j}_{b}(-k)e^{-i\vec{k}\cdot\vec{r}_{b}}
  \Big\}+o(m\alpha^8),
   \end{aligned} 
  \end{equation}  
where the total current in the electromagnetic field is   
  \begin{equation}
  \begin{aligned}\label{totalcurrent}
   J^{i}_{T}(k)e^{i\vec{k}\cdot\vec{r}}=&
   (J^{i}(k)+J^{i}_{ex}(k))e^{i\vec{k}\cdot\vec{r}}
   \\
  =&\dfrac{1}{2}\{\tilde{p},\sigma^{i}e^{i\vec{k}\cdot\vec{r}}\}
  -\dfrac{1}{8}
   \{p^{2},\{\tilde{p},\sigma^{i}e^{i\vec{k}\cdot\vec{r}}\}\}
  -\dfrac{1}{8}
  \omega[\tilde{p},\sigma^{i}e^{i\vec{k}\cdot\vec{r}}]
  \\&
  -\dfrac{1}{8}
  i[\sigma^ie^{i\vec{k}\cdot\vec{r}},\tilde{\mathscr{E}}(\vec{r})]
  +\mathscr{A}^{i}(\vec{r})e^{i\vec{k}\cdot\vec{r}}+o(m\alpha^3).
   \end{aligned} 
  \end{equation} 
   
The two retarded photons part of two photon-exchange interaction is 
 \begin{equation} 
  \begin{aligned}
  \Sigma_{2r}(E_{0})\equiv &\Sigma_{2prr}=
  \sum_{a,b,c,d,unequal}\int
  \dfrac{d_{ij}(k)d^{3}k}{2\omega_{k}(2\pi)^{3}}
  \dfrac{d_{kl}(q)d^{3}q}{2\omega_{q}(2\pi)^{3}}\times
  \\&
  \Big(
  J^{l}_{c}(q)e^{i\vec{q}\cdot\vec{r}_{c}}
    \tilde{G}(E_{0},\omega_{q})  
  J^{ik}_{a}(k,-q)e^{i(\vec{k}-\vec{q})\cdot\vec{r}_{a}}
    \tilde{G}(E_{0},\omega_{k})
  J^{j}_{b}(-k)e^{-i\vec{k}\cdot\vec{r}_{b}}+
  \\&
  J^{j}_{b}(k)e^{i\vec{k}\cdot\vec{r}_{b}}
  G(E_{0}-\omega_{k})
  J^{l}_{c}(q)e^{i\vec{q}\cdot\vec{r}_{c}}
  \tilde{G}(E_{0}-\omega_{k},\omega_{q})
  J^{ik}_{a}(-k,-q)e^{-i(\vec{k}+\vec{q})\cdot\vec{r}_{a}}+
  \\&
  J^{ik}_{a}(k,q)e^{i(\vec{k}+\vec{q})\cdot\vec{r}_{a}}
  \tilde{G}(E_{0}-\omega_{k},\omega_{q})  
  J^{j}_{b}(-k)e^{-i\vec{k}\cdot\vec{r}_{b}}
  G(E_{0}-\omega_{q})
  J^{l}_{c}(-q)e^{-i\vec{q}\cdot\vec{r}_{c}}+
  \\&
  J^{i}_{a}(k)e^{i\vec{k}\cdot\vec{r}_{a}}
  G(E_{0}-\omega_{k})
  J^{k}_{b}(q)e^{i\vec{q}\cdot\vec{r}_{a}}
  G(E_{0}-\omega_{k}-\omega_{q})
  J^{j}_{c}(-k)e^{-i\vec{k}\cdot\vec{r}_{c}}
  G(E_{0}-\omega_{q})
  J^{l}_{d}(-q)e^{-i\vec{q}\cdot\vec{r}_{d}}+
  \\&
  J^{i}_{a}(k)e^{i\vec{k}\cdot\vec{r}_{a}}
  G(E_{0}-\omega_{k})
  J^{k}_{b}(q)e^{i\vec{q}\cdot\vec{r}_{b}}
   \tilde{G}(E_{0}-\omega_{k},\omega_{q})
  J^{l}_{c}(-q)e^{-i\vec{q}\cdot\vec{r}_{c}}
  G(E_{0}-\omega_{k})
  J^{j}_{d}(-k)e^{-i\vec{k}\cdot\vec{r}_{d}}
  \Big).
   \end{aligned} 
  \end{equation}

The retarded photon-exchange interaction $\Sigma_{1r}$ and $\Sigma_{2r}$ perturb the energy-level. However, the corrections can't be obtained by using the perturbation theory. They are dependent on the energy of state.  The energy shift should be studied the pole of the total Green functions of the bound system   
 \begin{equation}
  \begin{aligned}
   G_{T}(E)=\dfrac{1}{E-H_{0}-\Sigma(E)},
  \end{aligned} 
 \end{equation}
where $\Sigma(E)$ is the relativistic corrections to photon-exchange interactions and relativistic kinetic energy.
 \begin{equation}
  \begin{aligned}
   \Sigma(E)
   \equiv H_{(4)}+H_{(6)}+H_{(8)}+\Sigma_{1r}(E)+\Sigma_{2r}(E)
   +o(m\alpha^8).
   \end{aligned} 
  \end{equation}   
The energy shift can be expressed as \citep{PhysRevA.71.012503}
 \begin{equation}
  \begin{aligned}
   \Delta E= \langle \Sigma(E_{0})\rangle
   +\langle \Sigma(E_{0})G'(E_{0})\Sigma(E_{0})\rangle
   +\langle \Sigma'(E_{0})\rangle
   \langle \Sigma(E_{0})\rangle+...,
  \end{aligned} 
 \end{equation}  
$\langle...\rangle$ is the mean value of the bound state. We can derive the Hamiltonian $H_{R1},H_{R2}$ of the photon retarded corrections by using this equation. The energy shifts are $\Delta E=\langle (H_{R1}+H_{R2})\rangle$

The Hamiltonian of the one photon retarded corrections is
  \begin{equation}\label{HR1}
  \begin{aligned}
  H_{R1}=&
  \sum_{a\neq b}\int 
  \dfrac{d_{ij}(\vec{k})d^{3}k}{2\omega(2\pi)^{3}}
  \Big\{J^{i}_{T(1),a}(k)e^{i\vec{k}\cdot\vec{r}_{a}}
  \tilde{G}(E_{0},\omega)
  J^{j}_{T(1),b}(-k)e^{-i\vec{k}\cdot\vec{r}_{b}}
  \\&
  +J^{i}_{T(1),a}(k)e^{i\vec{k}\cdot\vec{r}_{a}}
  \tilde{G}(E_{0},\omega)
  J^{j}_{T(3),b}(-k)e^{-i\vec{k}\cdot\vec{r}_{b}}
  +J^{i}_{T(3),a}(k)e^{i\vec{k}\cdot\vec{r}_{a}}
  \tilde{G}(E_{0},\omega)
  J^{j}_{T(1),b}(-k)e^{-i\vec{k}\cdot\vec{r}_{b}}
  \\&
  +J^{i}_{T(1)a}(k,q)e^{i\vec{k}\cdot\vec{r}_{a}}
  G'(E_{0}-\omega_{k})(H_{(4)}-\langle H_{(4)}\rangle)
  G'(E_{0}-\omega_{k})
  J^{j}_{T(1)b}(-k)e^{-i\vec{k}\cdot\vec{r}_{b}}
  \\&
  +H_{(4)}G'(E_{0})
  J^{i}_{T(1),a}(k)e^{i\vec{k}\cdot\vec{r}_{a}}
  \tilde{G}(E_{0},\omega)
  J^{j}_{T(1),b}(-k)e^{-i\vec{k}\cdot\vec{r}_{b}}
  \\&
  +J^{i}_{T(1),a}(k)e^{i\vec{k}\cdot\vec{r}_{a}}
  \tilde{G}(E_{0},\omega)
  J^{j}_{T(1),b}(-k)e^{-i\vec{k}\cdot\vec{r}_{b}}
  G'(E_{0})H_{(4)}
  \Big\}+o(m\alpha^8),
   \end{aligned} 
  \end{equation} 
where $G'$ is reduced Green function of fermion, and the total current (Eq.(\ref{totalcurrent})) in the electromagnetic field has been non-relativistic expanded as
  \begin{equation}
  \begin{aligned}
   &J^{i}_{T}(k)e^{i\vec{k}\cdot\vec{r}}=
   J^{i}_{T(1)}(k)e^{i\vec{k}\cdot\vec{r}}+
   J^{i}_{T(3)}(k)e^{i\vec{k}\cdot\vec{r}}+o(m\alpha^3),
   \\&
   J^{i}_{T(1)}(k)e^{i\vec{k}\cdot\vec{r}}=
   \dfrac{1}{2}\{\tilde{p},\sigma^{i}e^{i\vec{k}\cdot\vec{r}}\},
   \\&
   J^{i}_{T(3)}(k)e^{i\vec{k}\cdot\vec{r}}=
 -\dfrac{1}{8}
   \{p^{2},\{\tilde{p},\sigma^{i}e^{i\vec{k}\cdot\vec{r}}\}\}
  -\dfrac{1}{8}
  \omega[\tilde{p},\sigma^{i}e^{i\vec{k}\cdot\vec{r}}]
  -\dfrac{1}{8}
  i[\sigma^ie^{i\vec{k}\cdot\vec{r}},\tilde{\mathscr{E}}(\vec{r})]
  +\mathscr{A}^{i}(\vec{r})e^{i\vec{k}\cdot\vec{r}}.
   \end{aligned} 
  \end{equation} 
where the subscript $(n)$ means the $J^{i}_{T(n)}$ is at $m\alpha^n$ order. 

The first line in $H_{R1}$ is the non-relativistic part of one photon retarded corrections, which is at $m\alpha^5$ order. The order can be estimated by the following reason: the contributions of the energy of the virtual photon $\omega>m\alpha$ is suppressed in the $\tilde{G}(E_{0},\omega)$, the main contribution of the integral region is $\omega\sim m\alpha^2$. Another equivalent reason is the contribution of the virtual photon $\omega>m\alpha$ is the instantaneous interactions, which has been calculated with the non-retarded approximation (Eq.(\ref{potentialenergy})). The other terms in $H_{R1}$ are the relativistic corrections at $m\alpha^7$ order. As the relativistic effect will bring an $\alpha^2$ factor.     

The Hamiltonian of the two photon retarded correction is
   \begin{equation}\label{HR2}
  \begin{aligned}
 & H_{R2}=
  \sum_{a,b,c,d,unequal}\int
  \dfrac{d_{ij}(k)d^{3}k}{2\omega_{k}(2\pi)^{3}}
  \dfrac{d_{kl}(q)d^{3}q}{2\omega_{q}(2\pi)^{3}}\times
  \\&
  \Big(
  J^{l}_{c}(q)e^{i\vec{q}\cdot\vec{r}_{c}}
    \tilde{G}(E_{0},\omega_{q})  
  J^{ik}_{a}(k,-q)e^{i(\vec{k}-\vec{q})\cdot\vec{r}_{a}}
    \tilde{G}(E_{0},\omega_{k})
  J^{j}_{b}(-k)e^{-i\vec{k}\cdot\vec{r}_{b}}+
  \\&
  J^{j}_{b}(k)e^{i\vec{k}\cdot\vec{r}_{b}}
  G(E_{0}-\omega_{k})
  J^{l}_{c}(q)e^{i\vec{q}\cdot\vec{r}_{c}}
  \tilde{G}(E_{0}-\omega_{k},\omega_{q})
  J^{ik}_{a}(-k,-q)e^{-i(\vec{k}+\vec{q})\cdot\vec{r}_{a}}+
  \\&
  J^{ik}_{a}(k,q)e^{i(\vec{k}+\vec{q})\cdot\vec{r}_{a}}
  \tilde{G}(E_{0}-\omega_{k},\omega_{q})  
  J^{j}_{b}(-k)e^{-i\vec{k}\cdot\vec{r}_{b}}
  G(E_{0}-\omega_{q})
  J^{l}_{c}(-q)e^{-i\vec{q}\cdot\vec{r}_{c}}+
  \\&
  J^{i}_{a}(k)e^{i\vec{k}\cdot\vec{r}_{a}}
  G(E_{0}-\omega_{k})
  J^{k}_{b}(q)e^{i\vec{q}\cdot\vec{r}_{a}}
  G(E_{0}-\omega_{k}-\omega_{q})
  J^{j}_{c}(-k)e^{-i\vec{k}\cdot\vec{r}_{c}}
  G(E_{0}-\omega_{q})
  J^{l}_{d}(-q)e^{-i\vec{q}\cdot\vec{r}_{d}}+
  \\&
  J^{i}_{a}(k)e^{i\vec{k}\cdot\vec{r}_{a}}
  G(E_{0}-\omega_{k})
  J^{k}_{b}(q)e^{i\vec{q}\cdot\vec{r}_{b}}
   \tilde{G}(E_{0}-\omega_{k},\omega_{q})
  J^{l}_{c}(-q)e^{-i\vec{q}\cdot\vec{r}_{c}}
  G'(E_{0}-\omega_{k})
  J^{j}_{d}(-k)e^{-i\vec{k}\cdot\vec{r}_{d}}-
  \\&
  J^{i}_{a}(k)e^{i\vec{k}\cdot\vec{r}_{a}}
  G(E_{0}-\omega_{k})
  \langle
  J^{k}_{b}(q)e^{i\vec{q}\cdot\vec{r}_{b}}
   \tilde{G}(E_{0},\omega_{q})
  J^{l}_{c}(-q)e^{-i\vec{q}\cdot\vec{r}_{c}}
  \rangle
  G(E_{0}-\omega_{k})
  J^{j}_{d}(-k)e^{-i\vec{k}\cdot\vec{r}_{d}}
  \\&
  J^{i}_{a}(k)e^{i\vec{k}\cdot\vec{r}_{a}}
   \tilde{G}(E_{0},\omega_{k})
  J^{j}_{d}(-k)e^{-i\vec{k}\cdot\vec{r}_{d}}
  G'(E_{0})  
  J^{k}_{b}(q)e^{i\vec{q}\cdot\vec{r}_{b}}
   \tilde{G}(E_{0},\omega_{q})
  J^{l}_{c}(-q)e^{-i\vec{q}\cdot\vec{r}_{c}}
  \Big)+o(m\alpha^8),
   \end{aligned} 
  \end{equation}    
where $J^{ij}=\dfrac{1}{2}\delta^{ij}$ and $ J^{i}= J^{i}_{T(1)}$. As the main contribution of the integral region is $\omega_{k}\sim\omega_{q}\sim m\alpha^2$, which is the energy of the virtual photons, the leading order of $H_{R2}$ is $m\alpha^8$, and the relativistic correction to $H_{R2}$ is negligible up to $m\alpha^8$.

\section{SUMMATION}

In this work, we derived the Hamiltonian of NRQED by studying the one-photon-fermion and two-photon scattering matching in Sec.II. 
 \begin{equation}\label{HTotal}
  \begin{aligned}
    \mathscr{H}=& 
     \dfrac{1}{2}\tilde{\pi}^{2}+e\varphi
    -\dfrac{1}{8}\tilde{\pi}^{4}
    -\dfrac{i}{8}[\tilde{\pi},\tilde{E}]
     +\dfrac{1}{16}\tilde{\pi}^{6}
      +\dfrac{3i}{64}
     \{\tilde{\pi}^{2},[\tilde{\pi},\tilde{E}]\}
     +\dfrac{5i}{128}
     [\tilde{\pi}^{2},\{\tilde{\pi},\tilde{E}\}]
     +\dfrac{1}{8}\tilde{E}^{2} 
      \\&
     -\dfrac{5}{128}\tilde{\pi}^{8}
     -\dfrac{38i}{1024}
     [\tilde{\pi}^{4},\{\tilde{\pi},\tilde{E}\}]
     -\dfrac{40i}{1024}
     \{\tilde{\pi}^{4},[\tilde{\pi},\tilde{E}]\}
     +\dfrac{9i}{1024}
     [\tilde{\pi}^{2},[\tilde{\pi}^{2},[\tilde{\pi},\tilde{E}]]] 
     \\&
     -\dfrac{1}{64}
     \{\tilde{E},\{\tilde{\pi},\{\tilde{\pi},\tilde{E}\}\}\}
     -\dfrac{3}{64}
     \{\tilde{E},\{\tilde{\pi}^{2},\tilde{E}\}\}
     -\dfrac{i}{16}[\tilde{E},\partial_{t}\tilde{E}]
     \\&
     -\dfrac{5(1-x)}{64} \bigg\{
    i[\dfrac{\tilde{\pi}^{2}}{2},\{\tilde{\pi},\tilde{E}\}]
  +\{\tilde{\pi},\partial_{t}\tilde{E}\}+\{\tilde{E},\tilde{E}\}
  -i[\dfrac{\tilde{\pi}^{4}}{8},\{\tilde{\pi},\tilde{E}\}]
  -\dfrac{1}{8}
  \left[\{\tilde{\pi},\tilde{E}\},[\tilde{\pi},\tilde{E}]\right]
  \bigg\}.
   \end{aligned} 
  \end{equation}
We find the one-photon Hamiltonian has the different results by choosing different $x$, and the two-photon contact terms depend on the one-photon Hamiltonian terms. The anomalous magnetic moment corrections to the Hamiltonian are also obtained. 
 \begin{equation}
  \begin{aligned}
    \mathscr{H}_{AMM}=& 
     -\dfrac{i\kappa}{4}[\tilde{\pi},\tilde{E}]
     +\dfrac{i\kappa}{16}
     \{\tilde{\pi}^{2},[\tilde{\pi},\tilde{E}]\}
     +\dfrac{i\kappa}{32}
     [\tilde{\pi}^{2},\{\tilde{\pi},\tilde{E}\}]
     -\dfrac{\kappa}{2}\tilde{B}
     +\dfrac{\kappa}{16}[\tilde{\pi},[\tilde{\pi},\tilde{B}]]
     +\dfrac{1}{8}(2\kappa+\kappa^{2})\tilde{E}^{2},
   \end{aligned} 
  \end{equation}
where we have chosen the $x=1$. This Hamiltonian is coincided with the result of Pachucki up to $m\alpha^{6}$ order.  The $m\alpha^{8}$ order term is obtained first time.

The photon-exchange interactions are studied by using the NRQED Hamiltonian in the Sec.III. The interactions in non-retarded approximation are obtained, which is higher-order Breit interaction. It can be written as the potential energy between fermions (Eq.(\ref{potentialenergy})). The Hamiltonian Eq.(\ref{Hnr}) is sum of the kinetic energy and the potential energy, which is including the relativistic corrections. The result is coincided with other works at $m\alpha^{6}$ order. The $m\alpha^{8}$ order terms are also obtained. The higher-order energy shift of the non-retarded Hamiltonian can be calculated by using the perturbation theory. The retard photon-exchange interaction are studied in the second part of Sec.III. The one retarded photon contribution and two retarded photon contribution are obtained. These retarded corrections bring parts of Bethe logarithm terms in the multi-electron atoms. It is QED effect. We will study it in the future work. Additionally, If one of the fermion is heavier than others, The photon-exchange interaction can be expanded by the factor $m/M$. One can also obtained the recoil correction \citep{PhysRevA.94.052508,PhysRevA.95.012508}.

The photon can be emitted and absorbed by the same fermion. It is the self-energy corrections, which is divergence by using the NRQED Hamiltonian Eq.(\ref{HTotal}). The contribution of the high-energy virtual photon must be incorporated into the NRQED Hamiltonian. One has to do loop matching of the scattering amplitudes to obtain the contribution of the high-energy virtual photon. The infrared divergence in the high-energy region will be canceled with the ultraviolet divergence obtained by using the NRQED Hamiltonian \citep{PhysRevA.72.062102}.

In conclusion, the NRQED Hamiltonian can be using to calculate the relativistic, recoil and radiation corrections up to $m\alpha^{8}$ order in the future work.  

\newpage
\textbf{ACKNOWLEDGMENTS}

This work was supported by the National Natural Science Foundation of China (Grants No.11674253).

\bibliography{NRQED.bib}

\end{document}